\documentclass[prl,amsmath,amssymb,aps,superscriptaddress,floatfix,reprint,nofootinbib, notitlepage]{revtex4-1}
\usepackage[caption=false]{subfig}
\usepackage{caption}
\usepackage{ulem} 
\usepackage{float}
\usepackage{amsfonts,amssymb,amsmath}
\usepackage{graphics}
\usepackage{graphicx}
\usepackage{nicefrac}
\usepackage{cases}
\usepackage{mathrsfs}
\usepackage{enumerate}
\usepackage{mathtools}
\usepackage{textcomp}
\usepackage{verbatim}
\usepackage{bm}          
\usepackage{soul}
\usepackage{cancel}
\usepackage{upgreek}
\usepackage{txfonts}
\usepackage{color}
\usepackage[colorlinks=true, urlcolor=blue, linkcolor=blue]{hyperref}
\usepackage[papersize={8.5in,11in}]{geometry}
\usepackage{esint}
\renewcommand{\vec}[1]{\boldsymbol{#1}}
\newcommand {\be} {\begin{equation}}
\newcommand {\ee} {\end{equation}}

\usepackage{makecell}

\begin{document}

\title{Is disorder a friend or a foe to melting of Wigner-Mott insulators?}

\author{Mohammed Hammam}
\author{Cyprian Lewandowski}
\author{Vladimir Dobrosavljevic}
\author{Sandeep Joy}
\affiliation{National High Magnetic Field Laboratory, Tallahassee, Florida 32310, USA}
\affiliation{Department of Physics, Florida State University, Tallahassee, Florida 32306, USA}
\affiliation{FSU Quantum Initiative, Florida State University, Tallahassee, Florida 32306, USA}

\date{\today}
\begin{abstract}
Wigner crystals are extremely fragile, which is shown to result from very strong geometric frustration germane to long-range Coulomb interactions. Physically, this is manifested by a very small characteristic energy scale for shear density fluctuations, which are gapless excitations in a translationally invariant system. The presence of disorder, however, breaks translational invariance, thus suppressing gapless excitations and pushing them to higher density. We illustrate this general principle by explicit microscopic model calculations, showing that this mechanism very effectively stabilizes disordered Wigner lattices to much higher temperatures and densities than in the clean limit. On the other hand, we argue that in two dimensions disorder significantly ``smears" the melting transition, producing spatial coexistence of solid-like and liquid-like regions -- just as recently observed in STM experiments. Our results paint a new physical picture for melting of Wigner-Mott solids in two dimensions, corresponding to a Mott-Hubbard model with spatially varying local electronic bandwidth.  
\end{abstract}
\maketitle

At very low electron densities ($n$) and temperatures ($T$), a spatially uniform two-dimensional electron liquid spontaneously breaks translational symmetry and forms a solid arrangement known as a Wigner crystal (WC)~\cite{wigner1934on}. This phase transition arises from the competition between the interaction energy ($E_c\sim e^2 n^{1/2}/\epsilon_r$), which dominates over the kinetic energy (Here $-e<0$ is the electron charge and $\epsilon_r$ is the dielectric constant of the medium). The kinetic energy is either quantum mechanical ($E_F\sim\hbar^2 n/m$) or thermal ($\sim k_B T$), where $\hbar$ is the reduced Planck's constant, $m$ is the effective mass of the electron, and $k_B$ is the Boltzmann's constant. Although this phenomenon was proposed nearly a century ago, direct imaging of the WC phase, along with its melting transition, has only become possible recently~\cite{xiang2025imaging}\footnote{It must be noted that separate recent studies have imaged WC's stabilized either by periodic moir\'e modulation~\cite{Li2021imaging}—often termed 'generalized' WC's—or by strong magnetic fields that quench the kinetic energy~\cite{tsui2024wigner}.}.

Recent experimental observations, however, reveal notable discrepancies with existing theoretical predictions. Quantum Monte Carlo (QMC) simulations predict the melting of WC's at values of the dimensionless interaction parameter $r_s$ (defined as $r_s \equiv 1/(\pi n a_B^2)$, $a_B=\hbar^2\epsilon_r/m e^2$) around $\sim 30–34$~\cite{drummond2009phase, clark2009hexatic}\footnote{Hartree–Fock methods overestimate stability the crystalline phase and predict $r_s$ of the order unity; see Ref.~\cite{jain2025elementary} for a recent analysis.}. In contrast, experiments show that WC's remain stable at much lower $r_s$ values—that is, at significantly higher electron densities~\cite{huang2024electronic, joy2025disorder}. Moreover, the experiments in Ref.~\cite{xiang2025imaging, ge2025visualizing} report the existence of an intermediate phase in which both solid and liquid states coexist. Scenarios involving electronic microemulsions~\cite{spivak2003phase, spivak2004phases, jamei2005universal} and long-wavelength electron density fluctuations~\cite{shklovskii_completely_1972, ando1982electronic, shklovskii2007simple} have been ruled out as the origin of this observation. Instead, the enhanced stability and the intermediate phase have been attributed to short-range disorder based on the energetics argument~\cite{joy2025disorder}.

The melting of a WC can be understood in terms of the zero-point motion of the electrons that form the Wigner lattice: when the displacement amplitude becomes comparable to the lattice spacing, the crystal structure destabilizes and melts. This intuition lays the groundwork for the definition of the Lindemann ratio, an empirical criterion for predicting melting. In the clean limit, the Lindemann ratio has been widely used to compute phase diagrams and has proven successful in reproducing melting curves. It is defined as the ratio of the root-mean-square fluctuations around the equilibrium position to the lattice spacing. Remarkably, this ratio assumes a nearly universal value—typically between 0.20 and 0.25—across a wide range of melting transitions, whether classical or quantum, and regardless of whether the particles involved are bosons or fermions~\cite{AstrakharchikQuantum2007, babadi2013universal, kharpak2020lindemann}\footnote{The Lindemann ratio has been widely used in past studies of melting in the flux lattice of type-II superconductors~\cite{brandt1989thermal, houghton1989flux, nelson1989theory, fisher1991thermal}.}. This near-universal behavior of the Lindemann ratio is leveraged throughout this paper to characterize the phase diagram of the WC in the presence of disorder\footnote{The influence of impurities on electron solids and charge-density waves has been explored extensively in earlier theoretical studies \cite{fukuyam1978dynamics1, fukuyam1978dynamics2, fukuyama1978pinning, lee1979electric, normand1992pinning, ruzin1992pinning, cha1994orientational, cha1994topological, chitra1998dynamical, fertig1999electromagnetic, fogler2000dynamical, chitra2001pinned, chitra2005zero}, often inspired by analogous ideas developed for pinned vortex lattices in superconductors \cite{larkin1970effect, larkin1979pinning, blatter1994vortices}. However, this body of work has mainly concentrated on how impurity pinning modifies the static and dynamical properties.}.

\begin{figure*}[htb]
\centering
\includegraphics[width=2.0 \columnwidth]{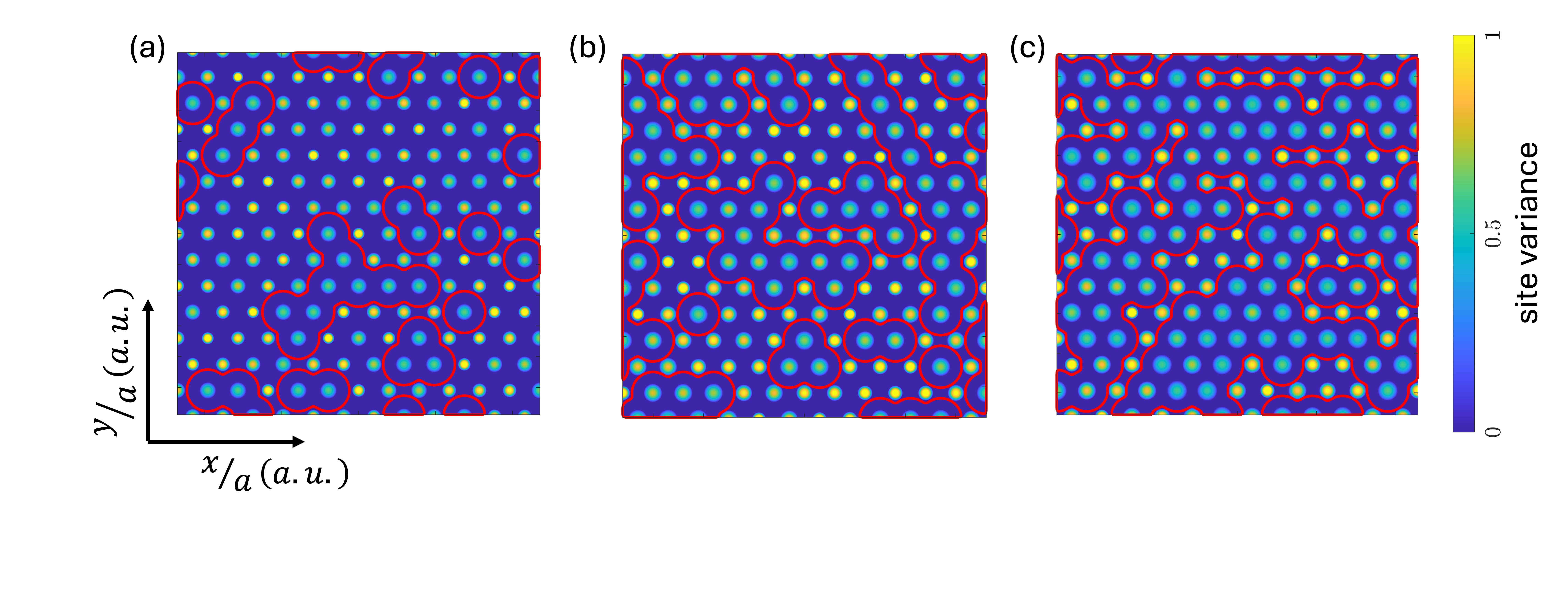}
\caption{This schematic figure illustrates how the fraction of melted sites increases as the electron density is tuned from low to high from left to right. The individual Gaussian wave packets are constructed using an Einstein-phonon approximation. Contour lines are drawn around regions that have locally melted; bright yellow denotes unmelted sites, whereas faded blue indicates melted sites. Here, the disorder strength used is $x_d=2\,\xi_0$, where $\xi_0$ is the bare bandwidth of the transverse phonon at the given density (see Ref.~\cite{SM}).} 
\label{fig: Melted-Real-Space}
\end{figure*}

Our main result can be summarized as follows. As expected, disorder stabilizes the WC phase to much higher densities. However, in the presence of disorder there is no sharp liquid–solid phase transition, but rather a crossover in which density fluctuations proliferate in certain regions (domains)\footnote{Formation of such domains has been indirectly inferred in the past from RF-conductivity experiments in magnetic field induced WC's; see, e.g., Refs.~\cite{moon2014pinning, freeman2024origin}.}. This is consistent with standard Imry–Ma arguments~\cite{imry1975random}: disorder explicitly breaks translational invariance and acts as a random field coupled to the crystalline order parameter. We characterize melting through a spatially varying vibrational amplitude. Physically, when this amplitude in a given region becomes comparable to the inter-particle spacing, particles there can exchange places and move around. Owing to disorder, the associated Lindemann ratios form a probability distribution, which we compute within our formalism. This picture can be visualized by placing a Gaussian wave packet on each site, with its width set by the local fluctuation amplitude. As temperature or density increases, some Gaussians broaden earlier and begin to overlap, forming connected liquid-like regions [illustrated in Fig.~\ref{fig: Melted-Real-Space}]. These real-space patterns closely resemble STM images in experiments and provide a compelling illustration of our theoretical framework.

In order to study the melting transition, let us set up the following low-energy effective Hamiltonian of a WC in the presence of impurities:
\begin{equation}
H_{\text{phonon}}=\sum_{i}\frac{\vec{p}_{i}^{2}}{2m}+\frac{1}{2}\sum_{i,j,\alpha,\beta}u_{i}^{\alpha}D_{ij}^{\alpha\beta}u_{j}^{\beta} +\frac{1}{2} \sum_{i} \kappa_{i}  \vec{u}_{i}^{2}.
\label{eq: H-phonon}
\end{equation}
Here, $\vec{u}_{i}$ denotes the displacement of the electron at site $i$ from its equilibrium position, and $D_{ij}^{\alpha\beta}$ is the dynamical matrix coupling fluctuations at sites $i$ and $j$ along directions $\alpha$ and $\beta$. The explicit form of $D_{ij}^{\alpha\beta}$ for a triangular WC is given in the supplemental material~\cite{SM}.
The last term accounts for disorder, with $\kappa_{i}$ representing the local onsite pinning potential. The physical intuition behind this disorder term is that, in the presence of atomic-scale, short-ranged defects—such as those relevant in STM experiments—the disorder length scale is much smaller than the spatial extent of the WC electron wave packet. Impurities that lie within a given wave packet effectively pin the electron locally, enhancing its confinement relative to the clean limit\footnote{Experiments in Refs.~\cite{xiang2025imaging, ge2025visualizing} also indicate the presence of long-ranged disorder arising from Coulomb impurities. Even in the presence of Coulomb disorder, an effective phonon model should still be possible. This might imply some degree of correlation between disorder on different sites, given the longer range of the Coulomb interaction, we leave an exploration of these effects to future work \cite{ruzin1992pinning}.}. We assume a Gaussian distribution for $\kappa_i$. As shown in the supplemental material, our results are qualitatively insensitive to the specific choice of disorder distribution, and we also derive the effective form of the disorder potential from first principles in the limit of weak, short-ranged disorder~\cite{SM}.

Here we first briefly discuss the physics of WC melting in the absence of disorder, before turning to the disorder‐dominated case (in the process, we will introduce the Lindemann ratio). The theory of two-dimensional crystals has a long history (see Ref.~\cite{strandburg1988two} for a review). Melting of a 2D WC, however, is more subtle than that of ordinary 2D solids, owing to the fermionic nature of electrons, their spin degrees of freedom, and the long-range Coulomb interaction. Nevertheless, the Lindemann ratio serves as a remarkably effective tool for understanding the melting of a WC~\cite{AstrakharchikQuantum2007, babadi2013universal, kharpak2020lindemann}. Its effectiveness is not accidental—although we do not attempt a first-principles derivation here. The key idea is that melting is governed by the collective Goldstone modes of the crystal. The dominant thermal or quantum fluctuations are the harmonic vibrational modes of the WC, with the shear mode setting the relevant Debye energy scale~\cite{thouless1978melting, fisher1982shear}. It is important to note that, since the crystal is two-dimensional, it supports two phonon modes: a transverse phonon with a linear dispersion, $\omega_{T,\vec{q}} \sim q$, and a longitudinal phonon with $\omega_{L,\vec{q}} \sim \sqrt{q}$\footnote{This is the plasmon mode that arises from the long-range nature of the Coulomb interactions and also exists in the liquid phase.}. Since the shear mode is far softer than the longitudinal mode for WC's, the melting is predominantly governed by transverse fluctuations. We have compared the melting boundary obtained by including both phonon modes with that obtained by considering only the transverse fluctuations and found that the two agree with high accuracy, as shown in the supplemental material~\cite{SM}. In the rest of the paper, we will restrict ourselves to discussing transverse phonons when examining the melting.

As discussed in the introduction, conventionally the Lindemann ratio is defined as $\eta^2_{\text{site}}\equiv \sigma^2_{\text{site}}/a^2$, where
\be
\sigma_{\text{site}}^{2}=\frac{1}{N}\sum_{i}\left\langle \vec{u}_{i}^{2}\right\rangle.    
\ee
Here, $i$ is the index of the site in the Wigner lattice in real space, and $N$ is the total number of sites. The subscript ``site" refers to the fact that this involves fluctuations of the sites, and it becomes clearer why we used that below. One can rewrite this in the momentum space in terms of the normal phonon modes as follows:
\be
\sigma_{\text{site}}^{2}=a_{c}\int\frac{d^2\vec{q}}{\left(2\pi\right)^{2}}\left(\frac{\hbar}{2m\omega_{\vec{q}}}\right)\coth\left[\frac{\hbar\omega_{\vec{q}}}{2k_{B}T}\right],
\label{eq: sigma-site}
\ee
where $a_c=\sqrt{3}a^2/2$ is the area of the unit cell of the Wigner lattice. $\sigma_{\text{site}}^{2}$ is a well defined quantity at $T=0$\footnote{At $T=0$, this is a $2+1$D quantum theory, which maps to a $3$D classical theory that is not subject to the curse of Mermin–Wagner theorem. In Eq.~(\ref{eq: sigma-site}), this corresponds to $\coth(\infty) = 1$, ensuring that the integral has no infrared divergence.}. However, at finite $T$, this momentum integral diverges logarithmically with the infrared cut-off at finite temperature $\sim T\log(1\big/q_{\text{min}}a)$, where $q_{\text{min}}$ is the lower cut-off of the integral, as dictated by the Mermin–Wagner theorem~\cite{kharpak2020lindemann}. This divergence prohibits us from using this to find the finite temperature melting point. One can alternatively define a generalized Lindemann ratio that measures relative (bond) fluctuations~\cite{lozovik1985oscillation, bedanov1985on, goldoni1996stability}. 
\be
 \sigma^{2}_{\text{bond}}\equiv\frac{1}{2NM}\sum_{i,j\in NN}\sum_{a}\left\langle\left(u_{i}^{a}-u_{j}^{a}\right)^{2}\right\rangle 
 \label{eq: sigma-bond}
 \ee
Here $M=6$ is the number of nearest neighbors. Equation~(\ref{eq: sigma-bond}) can be written in terms of the (shear) phonon modes as follows:
\be
\sigma_{\text{bond}}^{2}=\frac{a_{c}}{2 M}\int\frac{d^2\vec{q}}{\left(2\pi\right)^{2}}\left(\frac{\hbar}{2 m\omega_{\alpha,\vec{q}}}\right)\coth\left[\frac{\hbar\omega_{\alpha,\vec{q}}}{2k_{B}T}\right]f\left(\vec{q}\right)
\ee
Here $f(\vec{q})$ is the following geometrical factor:
\be
f\left(\vec{q}\right)=\sum_{j\in NN}4\sin^{2}\left(\frac{\vec{q}\cdot\vec{b}_{j}}{2}\right).
\ee
The vectors $\vec{b}_j$ point towards the nearest neighbor lattice sites. This approach effectively removes the logarithmic divergence by introducing a soft cut-off\footnote{In this sense, previous harmonic oscillator (HO) approaches~\cite{joy2022wigner, joy2023wigner, joy2025chiral} can be interpreted as focusing on relative fluctuations: it effectively pins all other electrons in place when computing the confining potential.} (in the limit $q\rightarrow0$, $f(\vec{q})$ can be expanded as, $\approx3q^2/4$). Following the literature where $\eta_\text{site}^{c}\approx0.23$~\cite{babadi2013universal}, corresponding $\eta_{\text{bond}}^{c}\approx0.25$ so that they reproduce the same zero temperature melting line. We note that this indicates the physics of 2D WC melting is controlled by the UV energy scale—specifically, the Debye energy or, equivalently, the large-$q$ contributions. This scale is the same in all dimensions, highlighting the distinctive nature of WC melting. This observation motivates our use of the Einstein-phonon approximation later for generating real-space charge-density plots.

\begin{figure*}[htb]
\centering
\includegraphics[width=2.0  \columnwidth]{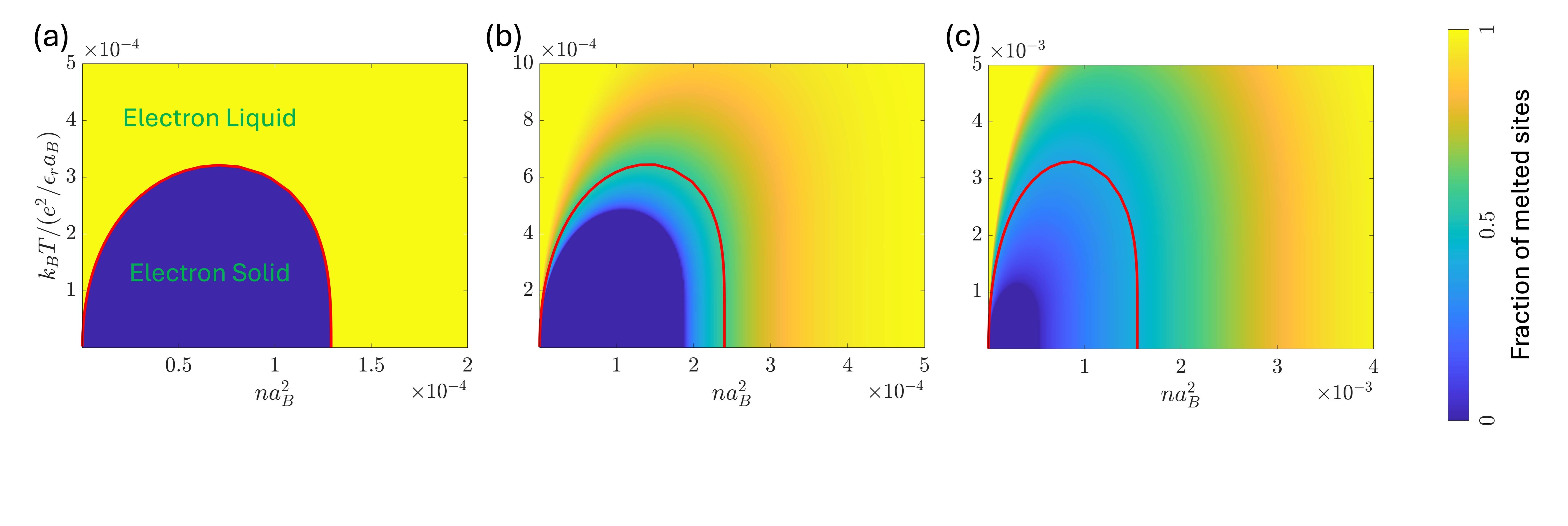}
\caption{The phase diagram of a two-dimensional electron system as a function of electron density and temperature for different disorder strengths. The solid red line denotes the melting line estimated from the global (bond) Lindemann ratio. The color gradient indicates the fraction of melted sites, calculated from the local (bond) Lindemann ratio. The color gradient goes from blue (dark shade) at zero to yellow (light shade) at one. The disorder strength is given in units of $\xi_0$, which is the bare bandwidth of the transverse phonon at the given density. (a) The clean limit is presented where there is ``nominally" a phase transition expected. (b) The disorder strength is given by $x_d=0.5\,\xi_0$ (c) The disorder strength is given by $x_d=2\,\xi_0$. As discussed in the main text, in the presence of disorder—which breaks translational symmetry—the melting/freezing transition that usually distinguishes a solid from a liquid is no longer well defined. Therefore, we do not label the phases in that way in the figures. (b) and (c).}
\label{fig: Melted-Fraction}
\end{figure*}

When dealing with disorder, it is more convenient to work in the language of Green’s functions. In the remainder of the paper, we therefore use the following retarded Green's function:\footnote{We are only considering the transverse fluctuations here.}:
\be
G_{j_{1}j_{2}}\left(t\right)\equiv-\dot{\iota}\theta\left(t\right)\left\langle \left[u_{j_{1}}\left(t\right),u_{j_{2}}\left(0\right)\right]\right\rangle
\ee
Here $\langle.\rangle$ denotes the finite temperature expectation value. In terms of the Green's function, relative bond fluctuations can be written as follows:
\be
 \sigma_{\text{bond}}^{2}=\frac{a_{c}}{2M}\sum_{a}\int\frac{d\vec{q}}{\left(2\pi\right)^{2}}\int_{-\infty}^{\infty}\frac{d\omega}{2\pi}n_{B}\left(\omega\right)B\left(\vec{q},\omega\right)f\left(\vec{q}\right)
\ee
where 
\be
B\left(\vec{q},\omega\right)\equiv-2\text{Im}\left(G\left(\vec{q},\omega\right)\right),
\ee
is the (bosonic) spectral function. We will also define a local Lindemann ratio as:
\be
 \sigma_{\text{bond-local}}^{2}=\frac{1}{2M}\sum_{i\in NN}\left\langle \left(\vec{u}_{0}-\vec{u}_{i}\right)^{2}\right\rangle 
\label{eq: sigma-bond-local}
\ee
In the clean limit, this quantity is equivalent to the spatially averaged quantity. However, in the case of a disordered situation, this quantity exhibits spatial fluctuations. We will be using Eq.~(\ref{eq: sigma-bond-local}) as a measure later to estimate the fraction of melted sites to construct the phase diagram. In terms of the Green's function language, this quantity is given by:
\be
\sigma_{\text{bond-local}}^{2}=\frac{1}{2}\int_{-\infty}^{\infty}\frac{d\omega}{2\pi}n_{B}\left(\omega\right)\left(B_{\kappa_{0}}\left(\omega\right)-\bar{B}\left(\omega\right)\right)+\sigma_{\text{bond}}^{2}
\label{eq: sigma-bond-local-green}
\ee
Here $B_{\kappa_{0}}\left(\omega\right)$ is the local spectral function as a function of the local pinning $\kappa_0$, and $\bar{B}(\omega)$ is the disorder-averaged local spectral function.

\begin{figure}[htb]
\centering
\includegraphics[width=1 \columnwidth]{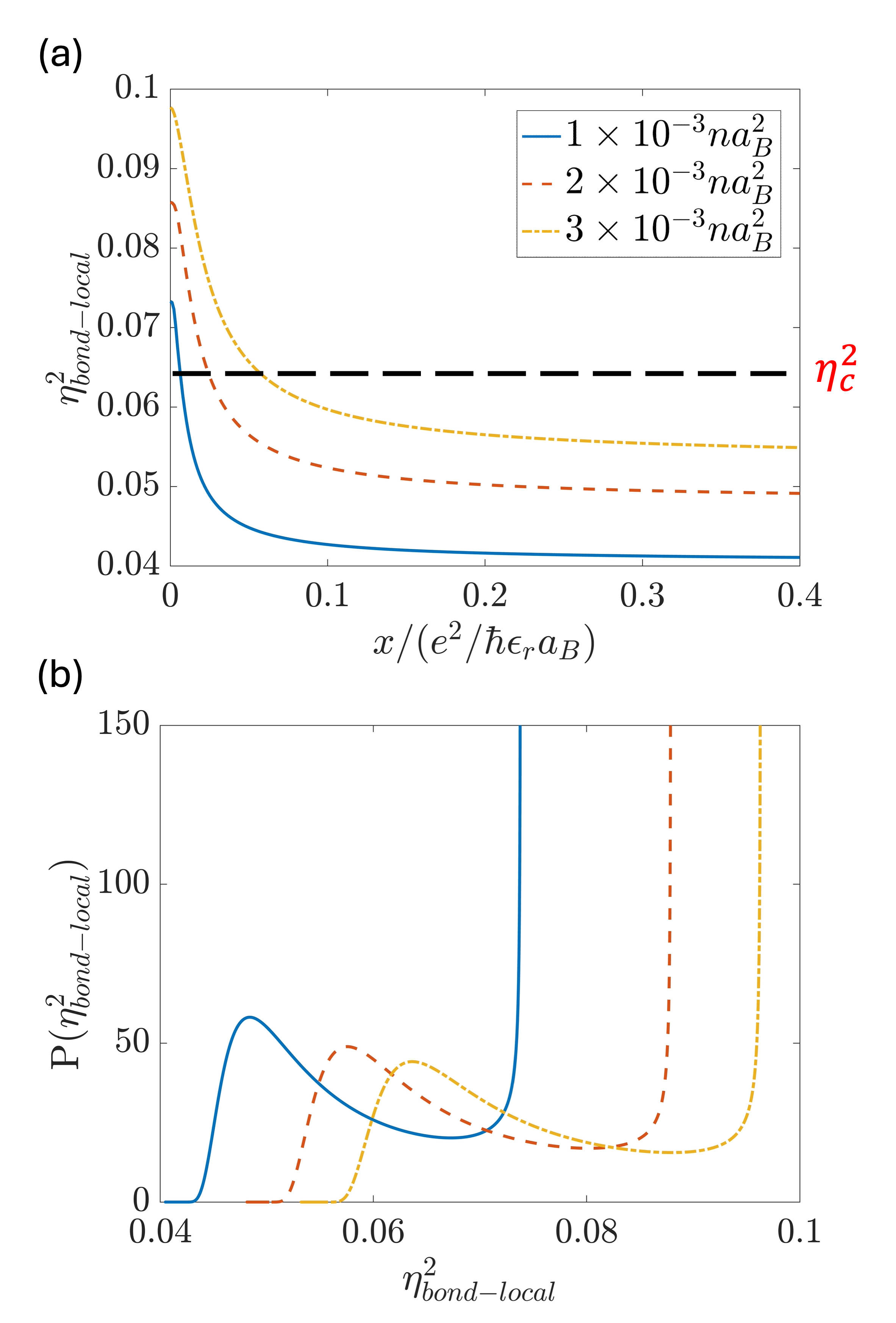}
\caption{(a) The local (bond) Lindemann ratio is plotted against the local disorder strength for three different densities at $x_d=2\,\xi_0$. The horizontal black dashed line marks the critical Lindemann ratio at which local melting occurs, implying that, depending on the local disorder strength, different lattice sites can undergo melting independently. (b) The probability distributions of the local bond Lindemann ratio for these same disorder strengths are numerically calculated and shown here.} 
\label{fig: gauss-prob-eta}
\end{figure}

We calculate the Green's function needed for the Lindemann ratio using dynamical mean-field theory within the coherent potential approximation (DMFT-CPA)~\cite{korringa1958dispersion, beeby1964electronic, soven1967coherent, georges1996dynamical, pankov2002pankov, ciuchi2018the}. This approach has been most commonly applied to study strong correlations and disorder in fermionic systems; however, its application to bosonic systems also exists (see, e.g., Refs.~\cite{Ghosh2002Phonons, Mondal2017Localization, mondal2019phonon}). For a bosonic system, the Green's function can be written as:
\be
G\left(z\right)=\int_{0}^{\xi_{0}}d\xi\frac{\rho_{0}\left(\xi\right)}{z^{2}-\Sigma\left(z\right)-\xi^{2}}.
\label{eq: G-CPA-SE}
\ee
Here  $z = \omega + i 0^+$  denotes the complex frequency, with $\omega$ as the real frequency and $ 0^+$ an infinitesimally small positive number ensuring causality. The integration variable $\xi$ denotes energy,  and $\rho_o(\xi)$ is the bare density of states (The details specific to the triangular WC are given in the supplemental material~\cite{SM}.). The quantity $\Sigma(z)$ is the self-consistent self-energy, which can be calculated locally within DMFT-CPA, and encodes the modification of the bare spectrum induced by disorder \footnote{$\Sigma(z)$ vanishes in the clean limit. In that case, Eq.~(\ref{eq: G-CPA-SE}) simply reproduces the real-frequency Green’s function obtained by Fourier transforming the momentum-space propagator.}. The next step in DMFT-CPA involves introducing a self-consistent cavity field $\Delta(z)$ which relates the Green's function to the distribution of the disorder as follows:
\be
G\left(z\right)=\int dx\frac{P\left(x\right)}{z^{2}-\Delta\left(z\right)-x^{2}}
\label{eq: G-CPA-cavity}
\ee
Here $x$ is related to the onsite pinning as $x=\sqrt{\kappa_i/m}$. One has to solve for the cavity field using the following relation:
\be
\Delta\left(z\right)=z^{2}-\Sigma\left(z\right)-G^{-1}\left(z\right)
\label{eq: G-CPA-SCE}
\ee
Details about the numerical techniques are provided in the supplemental material~\cite{SM}. 

Now that we have introduced the essential tools, we present our results. In Fig.~\ref{fig: Melted-Fraction}, we show the phase diagram of the WC in the density–temperature plane. Figure~\ref{fig: Melted-Fraction}(a) corresponds to the clean limit, where the Lindemann ratio indicates a sharp melting transition\footnote{Even in the clean limit, a direct first-order transition is replaced (possibly) by a series of intermediate electronic microemeulsion phases owing to the long-range nature of the interactions. However, the window of such intermediate phases are extremely narrow~\cite{joy2023upper}.}. The colors—blue (dark) to yellow (light)—represent the fraction of melted sites ranging from zero to one. As the strength of disorder increases, two apparent effects emerge. First, the solid line—representing the spatially averaged Lindemann ratio—shows that the crystal becomes stabilized to much higher densities, as seen in Fig.~\ref{fig: Melted-Fraction}(b) and Fig.~\ref{fig: Melted-Fraction}(c), corresponding to weak and strong disorder, respectively. Second, around this line, a broad region exists where the fraction of melted sites varies continuously from zero to one, indicating that there is no longer a sharp melting transition.

In Fig.~\ref{fig: gauss-prob-eta}(a) we demonstrate that the average local bond fluctuation decreases as the disorder strength increases. This quantity saturates at large $x$ due to the presence of the constant term in Eq.~(\ref{eq: sigma-bond-local-green}). Figure \ref{fig: gauss-prob-eta}(b) shows the probability distribution of the local bond Lindemann ratios. Each curve exhibits a (removable) singularity at the maximum allowed value of the local Lindemann ratio, which occurs at $x=0$. This behavior reflects the fact that $P(x)$ and $\eta^2_{\text{bond-local}}$ peaks at $x=0$\footnote{It is noteworthy that a local maximum emerges at finite $\eta^2_{\text{bond-local}}$. We attribute this to the convolution involved in mapping $P(x)$ to $P(\eta^2_{\text{bond-local}})$.}.

The absence of a sharp transition is further illustrated in Fig.~\ref{fig: Melted-Real-Space}, where we present real-space images. We construct a triangular lattice (used as an approximation to the triangulation observed experimentally) and sample local site fluctuations using the same Gaussian disorder as before. We have utilized the Einstein-phonon approximation to create this wave packet since $\langle u_i^2 \rangle$ nominally diverges at finite temperatures in the clean limit. The details of this construction are provided in the supplemental material~\cite{SM}. As the density increases (left to right), the number of sites satisfying the local Lindemann melting criterion grows, forming a percolating cluster or melted region—similar to what is seen in experiments. However, we emphasize that our theory, which relies on the Lindemann criterion, cannot reliably describe the melted regime. We can only assert that the system is no longer a solid; the nature of the melted phase (presumably a liquid with strong short-range correlations) is beyond the scope of our current description.

We conclude by saying that a frozen spatial charge profile--whether from STM or simulations--cannot sharply identify a ``melting" transition in terms of density (equivalently $r_s$) or temperatures. There are past experiments on the WC melting transition—using transport and optical spectroscopy—that report intermediate phases not consistent with either a conventional WC or a Fermi liquid (see e.g., Refs.~\cite{yoon1999wigner, knighton2018evidence, li2019evidence, hossain2020observation, falson2022competing, yang2023cascade, sung2025anelectronic}). These observations can be explained by the disorder-induced coexistence of clustered solid and liquid domains that we have found. However, there could be a well-defined transition: an insulator-metal transition (IMT), widely observed in two-dimensional electron systems as density is varied (pinned WC is an insulator, thus undergoes an IMT upon melting). We speculate that this IMT is Mott-like in nature. In the insulating phase, each WC site hosts a single localized electron forming a local moment, whereas the metallic phase is a disordered Fermi liquid~\cite{Camjayi2008coulomb, kravchenko2017strongly, tan2022how}. 

We note that after this work was completed, a related joint experiment–theory preprint~\cite{ge2025visualizing} appeared, using STM and QMC to investigate disorder effects on the WC. Their findings are consistent with our predictions.
\acknowledgments
\textit{Acknowledgments.} \; The authors are grateful to Ilya Esterlis, Zhehao Ge, Steve Kivelson, and Brian Skinner for useful discussions. S.J. acknowledges support from Florida State  University through the Quantum Postdoctoral Fellowship and the National High Magnetic Field Laboratory. C.L. was supported by start-up funds from Florida State University and the National High Magnetic Field Laboratory. M. H. and V. D. were supported by the NSF Grant No. DMR-2409911 and the National High Magnetic Field Laboratory. The National High Magnetic Field Laboratory is supported by the National Science Foundation through NSF/DMR-2128556 and the State of Florida.

\bibliography{ref}

@article{normand1992pinning,
  title = {Pinning and conductivity of a two-dimensional charge-density wave in a strong magnetic field},
  author = {Normand, B. G. A. and Littlewood, P. B. and Millis, A. J.},
  journal = {Phys. Rev. B},
  volume = {46},
  issue = {7},
  pages = {3920--3934},
  numpages = {0},
  year = {1992},
  month = {Aug},
  publisher = {American Physical Society},
  doi = {10.1103/PhysRevB.46.3920},
  url = {https://link.aps.org/doi/10.1103/PhysRevB.46.3920}
}

@article{ruzin1992pinning,
  title = {Pinning of a two-dimensional Wigner crystal by charged impurities},
  author = {Ruzin, I. M. and Marianer, S. and Shklovskii, B. I.},
  journal = {Phys. Rev. B},
  volume = {46},
  issue = {7},
  pages = {3999--4008},
  numpages = {0},
  year = {1992},
  month = {Aug},
  publisher = {American Physical Society},
  doi = {10.1103/PhysRevB.46.3999},
  url = {https://link.aps.org/doi/10.1103/PhysRevB.46.3999}
}

@article{fertig1999electromagnetic,
  title = {Electromagnetic response of a pinned Wigner crystal},
  author = {Fertig, H. A.},
  journal = {Phys. Rev. B},
  volume = {59},
  issue = {3},
  pages = {2120--2141},
  numpages = {0},
  year = {1999},
  month = {Jan},
  publisher = {American Physical Society},
  doi = {10.1103/PhysRevB.59.2120},
  url = {https://link.aps.org/doi/10.1103/PhysRevB.59.2120}
}

@article{fukuyama1978pinning,
  title = {Pinning and conductivity of two-dimensional charge-density waves in magnetic fields},
  author = {Fukuyama, Hidetoshi and Lee, Patrick A.},
  journal = {Phys. Rev. B},
  volume = {18},
  issue = {11},
  pages = {6245--6252},
  numpages = {0},
  year = {1978},
  month = {Dec},
  publisher = {American Physical Society},
  doi = {10.1103/PhysRevB.18.6245},
  url = {https://link.aps.org/doi/10.1103/PhysRevB.18.6245}
}

@article{fogler2000dynamical,
  title = {Dynamical response of a pinned two-dimensional Wigner crystal},
  author = {Fogler, Michael M. and Huse, David A.},
  journal = {Phys. Rev. B},
  volume = {62},
  issue = {11},
  pages = {7553--7570},
  numpages = {0},
  year = {2000},
  month = {Sep},
  publisher = {American Physical Society},
  doi = {10.1103/PhysRevB.62.7553},
  url = {https://link.aps.org/doi/10.1103/PhysRevB.62.7553}
}

@article{fukuyam1978dynamics1,
  title = {Dynamics of the charge-density wave. {I}. Impurity pinning in a single chain},
  author = {Fukuyama, H. and Lee, P. A.},
  journal = {Phys. Rev. B},
  volume = {17},
  issue = {2},
  pages = {535--541},
  numpages = {0},
  year = {1978},
  month = {Jan},
  publisher = {American Physical Society},
  doi = {10.1103/PhysRevB.17.535},
  url = {https://link.aps.org/doi/10.1103/PhysRevB.17.535}
}

@article{fukuyam1978dynamics2,
  title = {Dynamics of the charge-density wave. {II}. Long-range Coulomb effects in an array of chains},
  author = {Lee, P. A. and Fukuyama, H.},
  journal = {Phys. Rev. B},
  volume = {17},
  issue = {2},
  pages = {542--548},
  numpages = {0},
  year = {1978},
  month = {Jan},
  publisher = {American Physical Society},
  doi = {10.1103/PhysRevB.17.542},
  url = {https://link.aps.org/doi/10.1103/PhysRevB.17.542}
}

@article{bedanov1985on,
title = {On a modified Lindemann-like criterion for 2D melting},
journal = {Phys. Lett. A},
volume = {109},
number = {6},
pages = {289-291},
year = {1985},
issn = {0375-9601},
doi = {https://doi.org/10.1016/0375-9601(85)90617-6},
url = {https://www.sciencedirect.com/science/article/pii/0375960185906176},
author = {V.M. Bedanov and G.V. Gadiyak and Yu.E. Lozovik},
}

@article{goldoni1996stability,
  title = {Stability, dynamical properties, and melting of a classical bilayer Wigner crystal},
  author = {Goldoni, G. and Peeters, F. M.},
  journal = {Phys. Rev. B},
  volume = {53},
  issue = {8},
  pages = {4591--4603},
  numpages = {0},
  year = {1996},
  month = {Feb},
  publisher = {American Physical Society},
  doi = {10.1103/PhysRevB.53.4591},
  url = {https://link.aps.org/doi/10.1103/PhysRevB.53.4591}
}

@article{lozovik1985oscillation,
title = {Oscillation spectra and phase diagram of two-dimensional electron crystal: “New” (3+4)-self-consistent approximation},
journal = {Solid State Commun.},
volume = {54},
number = {8},
pages = {725-728},
year = {1985},
issn = {0038-1098},
doi = {https://doi.org/10.1016/0038-1098(85)90596-4},
url = {https://www.sciencedirect.com/science/article/pii/0038109885905964},
author = {Yu.E. Lozovik and V.M. Farztdinov},
}

@article{kharpak2020lindemann,
  title = {Lindemann melting criterion in two dimensions},
  author = {Khrapak, Sergey A.},
  journal = {Phys. Rev. Res.},
  volume = {2},
  issue = {1},
  pages = {012040},
  numpages = {6},
  year = {2020},
  month = {Feb},
  publisher = {American Physical Society},
  doi = {10.1103/PhysRevResearch.2.012040},
  url = {https://link.aps.org/doi/10.1103/PhysRevResearch.2.012040}
}

@article{joy2022wigner,
  title = {Wigner crystallization at large fine structure constant},
  author = {Joy, Sandeep and Skinner, Brian},
  journal = {Phys. Rev. B},
  volume = {106},
  issue = {4},
  pages = {L041402},
  numpages = {6},
  year = {2022},
  month = {Jul},
  publisher = {American Physical Society},
  doi = {10.1103/PhysRevB.106.L041402},
  url = {https://link.aps.org/doi/10.1103/PhysRevB.106.L041402}
}

@misc{joy2025disorder,
      title={Disorder-induced liquid-solid phase coexistence in {2D} electron systems}, 
      author={Sandeep Joy and Brian Skinner},
      year={2025},
      eprint={2502.11235},
      archivePrefix={arXiv},
}

@misc{jain2025elementary,
      title={Elementary Excitations, Melting Temperature and Correlation Energy in Wigner Crystal}, 
      author={Ambuj Jain and Chunli Huang},
      year={2025},
      eprint={2504.04538},
      archivePrefix={arXiv}
}

@article{xiang2025imaging,
author = {Ziyu Xiang  and Hongyuan Li  and Jianghan Xiao  and Mit H. Naik  and Zhehao Ge  and Zehao He  and Sudi Chen  and Jiahui Nie  and Shiyu Li  and Yifan Jiang  and Renee Sailus  and Rounak Banerjee  and Takashi Taniguchi  and Kenji Watanabe  and Sefaattin Tongay  and Steven G. Louie  and Michael F. Crommie  and Feng Wang },
title = {Imaging quantum melting in a disordered 2D Wigner solid},
journal = {Science},
volume = {388},
number = {6748},
pages = {736-740},
year = {2025},
doi = {10.1126/science.ado7136},
URL = {https://www.science.org/doi/abs/10.1126/science.ado7136},}

@article{soven1967coherent,
  title = {Coherent-Potential Model of Substitutional Disordered Alloys},
  author = {Soven, Paul},
  journal = {Phys. Rev.},
  volume = {156},
  issue = {3},
  pages = {809--813},
  numpages = {0},
  year = {1967},
  month = {Apr},
  publisher = {American Physical Society},
  doi = {10.1103/PhysRev.156.809},
  url = {https://link.aps.org/doi/10.1103/PhysRev.156.809}
}

@article{pankov2002pankov,
  title = {Semiclassical analysis of extended dynamical mean-field equations},
  author = {Pankov, Sergey and Kotliar, Gabriel and Motome, Yukitoshi},
  journal = {Phys. Rev. B},
  volume = {66},
  issue = {4},
  pages = {045117},
  numpages = {13},
  year = {2002},
  month = {Jul},
  publisher = {American Physical Society},
  doi = {10.1103/PhysRevB.66.045117},
  url = {https://link.aps.org/doi/10.1103/PhysRevB.66.045117}
}

@article{korringa1958dispersion,
title = {Dispersion theory for electrons in a random lattice with applications to the electronic structure of alloys},
journal = {J. Phys. Chem. Solids},
volume = {7},
number = {2},
pages = {252-258},
year = {1958},
issn = {0022-3697},
doi = {https://doi.org/10.1016/0022-3697(58)90270-1},
url = {https://www.sciencedirect.com/science/article/pii/0022369758902701},
author = {J. Korringa},
}

@article{beeby1964electronic,
  title = {Electronic Structure of Alloys},
  author = {Beeby, J. L.},
  journal = {Phys. Rev.},
  volume = {135},
  issue = {1A},
  pages = {A130--A143},
  numpages = {0},
  year = {1964},
  month = {Jul},
  publisher = {American Physical Society},
  doi = {10.1103/PhysRev.135.A130},
  url = {https://link.aps.org/doi/10.1103/PhysRev.135.A130}
}

@article{wigner1934on,
  title = {On the Interaction of Electrons in Metals},
  author = {Wigner, E.},
  journal = {Phys. Rev.},
  volume = {46},
  issue = {11},
  pages = {1002--1011},
  numpages = {0},
  year = {1934},
  month = {Dec},
  publisher = {American Physical Society},
  doi = {10.1103/PhysRev.46.1002},
  url = {https://link.aps.org/doi/10.1103/PhysRev.46.1002}
}

@Article{tsui2024wigner,
author={Tsui, Yen-Chen
and He, Minhao
and Hu, Yuwen
and Lake, Ethan
and Wang, Taige
and Watanabe, Kenji
and Taniguchi, Takashi
and Zaletel, Michael P.
and Yazdani, Ali},
title={Direct observation of a magnetic-field-induced Wigner crystal},
journal={Nature},
year={2024},
month={Apr},
day={01},
volume={628},
number={8007},
pages={287-292},
issn={1476-4687},
doi={10.1038/s41586-024-07212-7},
url={https://doi.org/10.1038/s41586-024-07212-7}
}

@article{drummond2009phase,
  title = {Phase Diagram of the Low-Density Two-Dimensional Homogeneous Electron Gas},
  author = {Drummond, N. D. and Needs, R. J.},
  journal = {Phys. Rev. Lett.},
  volume = {102},
  issue = {12},
  pages = {126402},
  numpages = {4},
  year = {2009},
  month = {Mar},
  publisher = {American Physical Society},
  doi = {10.1103/PhysRevLett.102.126402},
  url = {https://link.aps.org/doi/10.1103/PhysRevLett.102.126402}
}

@article{clark2009hexatic,
  title = {Hexatic and Mesoscopic Phases in a 2D Quantum Coulomb System},
  author = {Clark, Bryan K. and Casula, Michele and Ceperley, D. M.},
  journal = {Phys. Rev. Lett.},
  volume = {103},
  issue = {5},
  pages = {055701},
  numpages = {4},
  year = {2009},
  month = {Jul},
  publisher = {American Physical Society},
  doi = {10.1103/PhysRevLett.103.055701},
  url = {https://link.aps.org/doi/10.1103/PhysRevLett.103.055701}
}

@article{spivak2003phase,
  title = {Phase separation in the two-dimensional electron liquid in MOSFET's},
  author = {Spivak, B.},
  journal = {Phys. Rev. B},
  volume = {67},
  issue = {12},
  pages = {125205},
  numpages = {10},
  year = {2003},
  month = {Mar},
  publisher = {American Physical Society},
  doi = {10.1103/PhysRevB.67.125205},
  url = {https://link.aps.org/doi/10.1103/PhysRevB.67.125205}
}

@article{spivak2004phases,
  title = {Phases intermediate between a two-dimensional electron liquid and Wigner crystal},
  author = {Spivak, Boris and Kivelson, Steven A.},
  journal = {Phys. Rev. B},
  volume = {70},
  issue = {15},
  pages = {155114},
  numpages = {8},
  year = {2004},
  month = {Oct},
  publisher = {American Physical Society},
  doi = {10.1103/PhysRevB.70.155114},
  url = {https://link.aps.org/doi/10.1103/PhysRevB.70.155114}
}

@article{jamei2005universal,
  title = {Universal Aspects of Coulomb-Frustrated Phase Separation},
  author = {Jamei, Reza and Kivelson, Steven and Spivak, Boris},
  journal = {Phys. Rev. Lett.},
  volume = {94},
  issue = {5},
  pages = {056805},
  numpages = {4},
  year = {2005},
  month = {Feb},
  publisher = {American Physical Society},
  doi = {10.1103/PhysRevLett.94.056805},
  url = {https://link.aps.org/doi/10.1103/PhysRevLett.94.056805}
}

@article{shklovskii_completely_1972,
	title = {Completely compensated crystalline semiconductor as a model of an amorphous semiconductor},
	volume = {35},
	journal = {Sov. Phys.-JETP},
	author = {Shklovskii, B. I. and Efros, A. L.},
	year = {1972},
	pages = {610},
    url = {http://jetp.ras.ru/cgi-bin/dn/e_035_03_0610.pdf}
}

@article{ando1982electronic,
  title = {Electronic properties of two-dimensional systems},
  author = {Ando, Tsuneya and Fowler, Alan B. and Stern, Frank},
  journal = {Rev. Mod. Phys.},
  volume = {54},
  issue = {2},
  pages = {437--672},
  numpages = {0},
  year = {1982},
  month = {Apr},
  publisher = {American Physical Society},
  doi = {10.1103/RevModPhys.54.437},
  url = {https://link.aps.org/doi/10.1103/RevModPhys.54.437}
}

@article{shklovskii2007simple,
  title = {Simple model of Coulomb disorder and screening in graphene},
  author = {Shklovskii, B. I.},
  journal = {Phys. Rev. B},
  volume = {76},
  issue = {23},
  pages = {233411},
  numpages = {3},
  year = {2007},
  month = {Dec},
  publisher = {American Physical Society},
  doi = {10.1103/PhysRevB.76.233411},
  url = {https://link.aps.org/doi/10.1103/PhysRevB.76.233411}
}

@article{falson2022competing,
	title = {Competing correlated states around the zero-field {Wigner} crystallization transition of electrons in two dimensions},
	volume = {21},
	issn = {1476-4660},
	url = {https://doi.org/10.1038/s41563-021-01166-1},
	doi = {10.1038/s41563-021-01166-1},
	number = {3},
	journal = {Nature Materials},
	author = {Falson, J. and Sodemann, I. and Skinner, B. and Tabrea, D. and Kozuka, Y. and Tsukazaki, A. and Kawasaki, M. and von Klitzing, K. and Smet, J. H.},
	month = mar,
	year = {2022},
	pages = {311--316},
}

@misc{ge2025visualizing,
      title={Visualizing the Impact of Quenched Disorder on 2D Electron Wigner Solids}, 
      author={Zhehao Ge and Conor Smith and Zehao He and Yubo Yang and Qize Li and Ziyu Xiang and Jianghan Xiao and Wenjie Zhou and Salman Kahn and Melike Erdi and Rounak Banerjee and Takashi Taniguchi and Kenji Watanabe and Seth Ariel Tongay and Miguel A. Morales and Shiwei Zhang and Feng Wang and Michael F. Crommie},
      year={2025},
      eprint={2510.12009},
      archivePrefix={arXiv},
}

@article{babadi2013universal,
	doi = {10.1209/0295-5075/103/16002},
	url = {https://doi.org/10.1209/0295-5075/103/16002},
	year = 2013,
	month = {jul},
	publisher = {{IOP} Publishing},
	volume = {103},
	number = {1},
	pages = {16002},
	author = {Mehrtash Babadi and Brian Skinner and Michael M. Fogler and Eugene Demler},
	title = {Universal behavior of repulsive two-dimensional fermions in the vicinity of the quantum freezing point},
	journal = {{EPL} (Europhysics Letters)}
}

@article{AstrakharchikQuantum2007,
  title = {Quantum Phase Transition in a Two-Dimensional System of Dipoles},
  author = {Astrakharchik, G. E. and Boronat, J. and Kurbakov, I. L. and Lozovik, Yu. E.},
  journal = {Phys. Rev. Lett.},
  volume = {98},
  issue = {6},
  pages = {060405},
  numpages = {4},
  year = {2007},
  month = {Feb},
  publisher = {American Physical Society},
  doi = {10.1103/PhysRevLett.98.060405},
  url = {https://link.aps.org/doi/10.1103/PhysRevLett.98.060405}
}

@misc{joy2023wigner,
      title={Wigner crystallization in Bernal bilayer graphene}, 
      author={Sandeep Joy and Brian Skinner},
      year={2023},
      eprint={2310.07751},
      archivePrefix={arXiv},
}

@note{SM,
note={Supplemental materials will be uploaded soon.}
}

@article{strandburg1988two,
  title = {Two-dimensional melting},
  author = {Strandburg, Katherine J.},
  journal = {Rev. Mod. Phys.},
  volume = {60},
  issue = {1},
  pages = {161--207},
  numpages = {0},
  year = {1988},
  month = {Jan},
  publisher = {American Physical Society},
  doi = {10.1103/RevModPhys.60.161},
  url = {https://link.aps.org/doi/10.1103/RevModPhys.60.161}
}

@article{thouless1978melting,
doi = {10.1088/0022-3719/11/6/001},
url = {https://doi.org/10.1088/0022-3719/11/6/001},
year = {1978},
month = {mar},
publisher = {},
volume = {11},
number = {6},
pages = {L189},
author = {D J Thouless},
title = {Melting of the two-dimensional {W}igner lattice},
journal = {J. Phys. C: Solid State Phys.}
}

@article{fisher1982shear,
  title = {Shear moduli and melting temperatures of two-dimensional electron crystals: Low temperatures and high magnetic fields},
  author = {Fisher, Daniel S.},
  journal = {Phys. Rev. B},
  volume = {26},
  issue = {9},
  pages = {5009--5021},
  numpages = {0},
  year = {1982},
  month = {Nov},
  publisher = {American Physical Society},
  doi = {10.1103/PhysRevB.26.5009},
  url = {https://link.aps.org/doi/10.1103/PhysRevB.26.5009}
}

@article{joy2025chiral,
      title={Chiral Wigner crystal phases induced by Berry curvature}, 
      author={Sandeep Joy and Leonid Levitov and Brian Skinner},
      eprint={2507.22121},
      archivePrefix={arXiv},
}

@article{georges1996dynamical,
  title = {Dynamical mean-field theory of strongly correlated fermion systems and the limit of infinite dimensions},
  author = {Georges, Antoine and Kotliar, Gabriel and Krauth, Werner and Rozenberg, Marcelo J.},
  journal = {Rev. Mod. Phys.},
  volume = {68},
  issue = {1},
  pages = {13--125},
  numpages = {0},
  year = {1996},
  month = {Jan},
  publisher = {American Physical Society},
  doi = {10.1103/RevModPhys.68.13},
  url = {https://link.aps.org/doi/10.1103/RevModPhys.68.13}
}

@article{Ghosh2002Phonons,
  title = {Phonons in random alloys: The itinerant coherent-potential approximation},
  author = {Ghosh, Subhradip and Leath, P. L. and Cohen, Morrel H.},
  journal = {Phys. Rev. B},
  volume = {66},
  issue = {21},
  pages = {214206},
  numpages = {16},
  year = {2002},
  month = {Dec},
  publisher = {American Physical Society},
  doi = {10.1103/PhysRevB.66.214206},
  url = {https://link.aps.org/doi/10.1103/PhysRevB.66.214206}
}

@Article{Camjayi2008coulomb,
author={Camjayi, A.
and Haule, K.
and Dobrosavljevi{\'{c}}, V.
and Kotliar, G.},
title={Coulomb correlations and the Wigner--Mott transition},
journal={Nature Physics},
year={2008},
month={Dec},
day={01},
volume={4},
number={12},
pages={932-935},
issn={1745-2481},
doi={10.1038/nphys1106},
url={https://doi.org/10.1038/nphys1106}
}

@book{kravchenko2017strongly,
  title={Strongly Correlated Electrons in Two Dimensions},
  author={Kravchenko, S.},
  isbn={9789814745383},
  year={2017},
  publisher={Jenny Stanford Publishing}
}

@Article{tan2022how,
AUTHOR = {Tan, Yuting and Dobrosavljević, Vladimir and Rademaker, Louk},
TITLE = {How to Recognize the Universal Aspects of Mott Criticality?},
JOURNAL = {Crystals},
VOLUME = {12},
YEAR = {2022},
NUMBER = {7},
ARTICLE-NUMBER = {932},
ISSN = {2073-4352},
DOI = {10.3390/cryst12070932}
}

@article{Mondal2017Localization,
  title = {Localization of phonons in mass-disordered alloys: A typical medium dynamical cluster approach},
  author = {Mondal, Wasim Raja and Vidhyadhiraja, N. S. and Berlijn, T. and Moreno, Juana and Jarrell, M.},
  journal = {Phys. Rev. B},
  volume = {96},
  issue = {1},
  pages = {014203},
  numpages = {12},
  year = {2017},
  month = {Jul},
  publisher = {American Physical Society},
  doi = {10.1103/PhysRevB.96.014203},
  url = {https://link.aps.org/doi/10.1103/PhysRevB.96.014203}
}

@article{joy2023upper,
  title = {Upper bound on the window of density occupied by microemulsion phases in two-dimensional electron systems},
  author = {Joy, Sandeep and Skinner, Brian},
  journal = {Phys. Rev. B},
  volume = {108},
  issue = {24},
  pages = {L241110},
  numpages = {6},
  year = {2023},
  month = {Dec},
  publisher = {American Physical Society},
  doi = {10.1103/PhysRevB.108.L241110},
  url = {https://link.aps.org/doi/10.1103/PhysRevB.108.L241110}
}

@Article{ciuchi2018the,
author={Ciuchi, Sergio
and Di Sante, Domenico
and Dobrosavljevi{\'{c}}, Vladimir
and Fratini, Simone},
title={The origin of Mooij correlations in disordered metals},
journal={npj Quantum Materials},
year={2018},
month={Sep},
day={13},
volume={3},
number={1},
pages={44},
issn={2397-4648},
doi={10.1038/s41535-018-0119-y},
url={https://doi.org/10.1038/s41535-018-0119-y}
}

@article{huang2024electronic,
  title = {Electronic transport, metal-insulator transition, and Wigner crystallization in transition metal dichalcogenide monolayers},
  author = {Huang, Yi and Das Sarma, Sankar},
  journal = {Phys. Rev. B},
  volume = {109},
  issue = {24},
  pages = {245431},
  numpages = {26},
  year = {2024},
  month = {Jun},
  publisher = {American Physical Society},
  doi = {10.1103/PhysRevB.109.245431},
  url = {https://link.aps.org/doi/10.1103/PhysRevB.109.245431}
}

@Article{Li2021imaging,
author={Li, Hongyuan
and Li, Shaowei
and Regan, Emma C.
and Wang, Danqing
and Zhao, Wenyu
and Kahn, Salman
and Yumigeta, Kentaro
and Blei, Mark
and Taniguchi, Takashi
and Watanabe, Kenji
and Tongay, Sefaattin
and Zettl, Alex
and Crommie, Michael F.
and Wang, Feng},
title={Imaging two-dimensional generalized Wigner crystals},
journal={Nature},
year={2021},
month={Sep},
day={01},
volume={597},
number={7878},
pages={650-654},
doi={10.1038/s41586-021-03874-9},
url={https://doi.org/10.1038/s41586-021-03874-9}
}

@article{moon2014pinning,
  title = {Pinning modes of high-magnetic-field Wigner solids with controlled alloy disorder},
  author = {Moon, B.-H. and Engel, L. W. and Tsui, D. C. and Pfeiffer, L. N. and West, K. W.},
  journal = {Phys. Rev. B},
  volume = {89},
  issue = {7},
  pages = {075310},
  numpages = {5},
  year = {2014},
  month = {Feb},
  publisher = {American Physical Society},
  doi = {10.1103/PhysRevB.89.075310},
  url = {https://link.aps.org/doi/10.1103/PhysRevB.89.075310}
}

@article{freeman2024origin,
  title = {Origin of Pinning Disorder in Magnetic-Field-Induced Wigner Solids},
  author = {Freeman, Matthew L. and Madathil, P. T. and Pfeiffer, L. N. and Baldwin, K. W. and Chung, Y. J. and Winkler, R. and Shayegan, M. and Engel, L. W.},
  journal = {Phys. Rev. Lett.},
  volume = {132},
  issue = {17},
  pages = {176301},
  numpages = {6},
  year = {2024},
  month = {Apr},
  publisher = {American Physical Society},
  doi = {10.1103/PhysRevLett.132.176301},
  url = {https://link.aps.org/doi/10.1103/PhysRevLett.132.176301}
}

@article{imry1975random,
  title = {Random-Field Instability of the Ordered State of Continuous Symmetry},
  author = {Imry, Yoseph and Ma, Shang Keng},
  journal = {Phys. Rev. Lett.},
  volume = {35},
  issue = {21},
  pages = {1399--1401},
  numpages = {0},
  year = {1975},
  month = {Nov},
  publisher = {American Physical Society},
  doi = {10.1103/PhysRevLett.35.1399},
  url = {https://link.aps.org/doi/10.1103/PhysRevLett.35.1399}
}

@article{mondal2019phonon,
  title = {Phonon localization in binary alloys with diagonal and off-diagonal disorder: A cluster Green's function approach},
  author = {Mondal, Wasim Raja and Berlijn, T. and Jarrell, M. and Vidhyadhiraja, N. S.},
  journal = {Phys. Rev. B},
  volume = {99},
  issue = {13},
  pages = {134203},
  numpages = {11},
  year = {2019},
  month = {Apr},
  publisher = {American Physical Society},
  doi = {10.1103/PhysRevB.99.134203},
  url = {https://link.aps.org/doi/10.1103/PhysRevB.99.134203}
}

@article{brandt1989thermal,
  title = {Thermal fluctuation and melting of the vortex lattice in oxide superconductors},
  author = {Brandt, E. H.},
  journal = {Phys. Rev. Lett.},
  volume = {63},
  issue = {10},
  pages = {1106--1109},
  numpages = {0},
  year = {1989},
  month = {Sep},
  publisher = {American Physical Society},
  doi = {10.1103/PhysRevLett.63.1106},
  url = {https://link.aps.org/doi/10.1103/PhysRevLett.63.1106}
}

@article{houghton1989flux,
  title = {Flux lattice melting in high-${T}_{c}$ superconductors},
  author = {Houghton, A. and Pelcovits, R. A. and Sudb\o{}, A.},
  journal = {Phys. Rev. B},
  volume = {40},
  issue = {10},
  pages = {6763--6770},
  numpages = {0},
  year = {1989},
  month = {Oct},
  publisher = {American Physical Society},
  doi = {10.1103/PhysRevB.40.6763},
  url = {https://link.aps.org/doi/10.1103/PhysRevB.40.6763}
}

@article{nelson1989theory,
  title = {Theory of melted flux liquids},
  author = {Nelson, David R. and Seung, H. Sebastian},
  journal = {Phys. Rev. B},
  volume = {39},
  issue = {13},
  pages = {9153--9174},
  numpages = {0},
  year = {1989},
  month = {May},
  publisher = {American Physical Society},
  doi = {10.1103/PhysRevB.39.9153},
  url = {https://link.aps.org/doi/10.1103/PhysRevB.39.9153}
}

@article{fisher1991thermal,
  title = {Thermal fluctuations, quenched disorder, phase transitions, and transport in type-II superconductors},
  author = {Fisher, Daniel S. and Fisher, Matthew P. A. and Huse, David A.},
  journal = {Phys. Rev. B},
  volume = {43},
  issue = {1},
  pages = {130--159},
  numpages = {0},
  year = {1991},
  month = {Jan},
  publisher = {American Physical Society},
  doi = {10.1103/PhysRevB.43.130},
  url = {https://link.aps.org/doi/10.1103/PhysRevB.43.130}
}

@article{knighton2018evidence,
  title = {Evidence of two-stage melting of Wigner solids},
  author = {Knighton, Talbot and Wu, Zhe and Huang, Jian and Serafin, Alessandro and Xia, J. S. and Pfeiffer, L. N. and West, K. W.},
  journal = {Phys. Rev. B},
  volume = {97},
  issue = {8},
  pages = {085135},
  numpages = {6},
  year = {2018},
  month = {Feb},
  publisher = {American Physical Society},
  doi = {10.1103/PhysRevB.97.085135},
  url = {https://link.aps.org/doi/10.1103/PhysRevB.97.085135}
}

@article{yoon1999wigner,
  title = {Wigner Crystallization and Metal-Insulator Transition of Two-Dimensional Holes in GaAs at $\mathit{B}\phantom{\rule{0ex}{0ex}}=\phantom{\rule{0ex}{0ex}}0$},
  author = {Yoon, Jongsoo and Li, C. C. and Shahar, D. and Tsui, D. C. and Shayegan, M.},
  journal = {Phys. Rev. Lett.},
  volume = {82},
  issue = {8},
  pages = {1744--1747},
  numpages = {0},
  year = {1999},
  month = {Feb},
  publisher = {American Physical Society},
  doi = {10.1103/PhysRevLett.82.1744},
  url = {https://link.aps.org/doi/10.1103/PhysRevLett.82.1744}
}

@article{hossain2020observation,
author = {M. S. Hossain  and M. K. Ma  and K. A. Villegas Rosales  and Y. J. Chung  and L. N. Pfeiffer  and K. W. West  and K. W. Baldwin  and M. Shayegan },
title = {Observation of spontaneous ferromagnetism in a two-dimensional electron system},
journal = {PNAS},
volume = {117},
number = {51},
pages = {32244-32250},
year = {2020},
doi = {10.1073/pnas.2018248117},
URL = {https://www.pnas.org/doi/abs/10.1073/pnas.2018248117}}

@article{sung2025anelectronic,
author = {Sung, Jiho and Wang, Jue and Esterlis, Ilya and Volkov, Pavel A and Scuri, Giovanni and Zhou, You and Brutschea, Elise and Taniguchi, Takashi and Watanabe, Kenji and Yang, Yubo and Morales, Miguel A and Zhang, Shiwei and Millis, Andrew J and Lukin, Mikhail D and Kim, Philip and Demler, Eugene and Park, Hongkun},
issn = {1745-2481},
journal = {Nature Physics},
title = {{An electronic microemulsion phase emerging from a quantum crystal-to-liquid transition}},
url = {https://doi.org/10.1038/s41567-024-02759-8},
year = {2025}
}

@article{li2019evidence,
  title = {Evidence for mixed phases and percolation at the metal-insulator transition in two dimensions},
  author = {Li, Shiqi and Zhang, Qing and Ghaemi, Pouyan and Sarachik, M. P.},
  journal = {Phys. Rev. B},
  volume = {99},
  issue = {15},
  pages = {155302},
  numpages = {7},
  year = {2019},
  month = {Apr},
  publisher = {American Physical Society},
  doi = {10.1103/PhysRevB.99.155302},
  url = {https://link.aps.org/doi/10.1103/PhysRevB.99.155302}
}

@article{yang2023cascade,
  title = {Cascade of Multielectron Bubble Phases in Monolayer Graphene at High Landau Level Filling},
  author = {Yang, Fangyuan and Bai, Ruiheng and Zibrov, Alexander A. and Joy, Sandeep and Taniguchi, Takashi and Watanabe, Kenji and Skinner, Brian and Goerbig, Mark O. and Young, Andrea F.},
  journal = {Phys. Rev. Lett.},
  volume = {131},
  issue = {22},
  pages = {226501},
  numpages = {5},
  year = {2023},
  month = {Nov},
  publisher = {American Physical Society},
  doi = {10.1103/PhysRevLett.131.226501},
  url = {https://link.aps.org/doi/10.1103/PhysRevLett.131.226501}
}

@article{cha1994topological,
  title = {Topological defects, orientational order, and depinning of the electron solid in a random potential},
  author = {Cha, Min-Chul and Fertig, H. A.},
  journal = {Phys. Rev. B},
  volume = {50},
  issue = {19},
  pages = {14368--14380},
  numpages = {0},
  year = {1994},
  month = {Nov},
  publisher = {American Physical Society},
  doi = {10.1103/PhysRevB.50.14368},
  url = {https://link.aps.org/doi/10.1103/PhysRevB.50.14368}
}

\end{document}